\documentstyle[epsf]{aipproc}
\def\V{Vela X--1}   \def\B{BeppoSAX}
\begin{document}

\title{\B\ observation of the X--ray binary pulsar \V}
\author{M. Orlandini$^1$, D. Dal Fiume$^1$, L. Nicastro$^1$, \\
        S. Giarrusso$^2$, A. Segreto$^2$, S. Piraino$^2$, \\
        G. Cusumano$^2$, S. Del Sordo$^2$,
        M. Guainazzi$^3$,
        L. Piro$^4$}
\address{$^1$ Istituto TESRE/CNR, via Gobetti 101, 40129 Bologna, Italy \\
 $^2$ Istituto IFCAI/CNR, via La~Malfa 153, 90146 Palermo, Italy \\
 $^3$ SAX/Scientific Data Center/Nuova Telespazio, via Corcolle 19, 00131 Rome, Italy \\
 $^4$ Istituto Astrofisica Spaziale/CNR, via Fermi 21, 00044 Frascati, Italy}

\maketitle

\begin{abstract}
We report on the spectral (pulse averaged) and timing analysis of the $\sim 20$
ksec observation of the X--ray binary pulsar Vela X--1 performed during the \B\
Science Verification Phase. The source was observed in two different intensity
states: the low state is probably due to an erratic intensity dip and shows a
decrease of a factor $\sim 2$ in intensity, and a factor 10 in $N_H$. We have
not been able to fit the 2--100 keV continuum spectrum with the standard (for
an X--ray pulsar) power law modified by a high energy cutoff because of the
flattening of the spectrum in $\sim 10$--30 keV. The timing analysis confirms
previous results: the pulse profile changes from a five-peak structure for
energies less than 15 keV, to a simpler two-peak shape at higher energies. The
Fourier analysis shows a very complex harmonic component: up to 23 harmonics
are clearly visible in the power spectrum, with a dominant first harmonic for
low energy data, and a second one as the more prominent for energies greater
than 15 keV. The aperiodic component in the Vela X--1 power spectrum presents a
{\em knee\/} at about 1 Hz. The pulse period, corrected for binary motion, is
$283.206 \pm 0.001$ sec.
\end{abstract}

\section*{INTRODUCTION}

\V\ is the prototype of the class of accreting high-mass X--ray binary pulsars,
with a spectrum that was fit by a power law modified at high energy by an
exponential cutoff \cite{303}. Moreover, the spectrum shows line features: a
$\sim 6.4$ keV Iron fluorescence emission line, and a cyclotron resonance
feature (CRF) at $\sim 55$ keV \cite{1581}. The pulse-averaged spectrum depends
on the 8.96 day orbital period \cite{1275}: episodes of strong absorption are
explained in terms of accretion \cite{88} or photoionization \cite{1345} wakes.

The pulse period history of the 283 sec X--ray binary pulsar \V\ shows a
typical wavy behavior \cite{1174}. The reversal of spin-up and spin-down has
been explained in term of a (temporary) accretion disk, acting as reservoir for
the momentum transfer \cite{1333}.

\section*{OBSERVATION}

\B\ is a program of the Italian Space Agency (ASI) with participation of the
Netherlands Agency for Aerospace Programs (NIVR). It is composed by four
co-aligned Narrow Field Instruments (NFIs) \cite{1530}, operating in the energy
ranges 0.1--10 keV (LECS, not operative during this observation) \cite{1531},
1--10 keV (MECS) \cite{1532}, 3--120 keV (HPGSPC) \cite{1533} and 15--300 keV
(PDS) \cite{1386}.  Perpendicular to the NFI axis there are two Wide Field
Cameras \cite{1534}, with a $40^\circ \times 40^\circ$ field of view.

During the \B\ Science Verification Phase (SVP) \V\ was observed on 1996 July
14 from 06:01 to 20:54 UT, corresponding to orbital phases 0.28--0.35
\cite{705}. All the instruments were operated in direct modes, which provide
information on each photon. The two mechanical-collimated instruments were
operated in rocking mode with a 96 sec stay time for HPGSPC and 50 sec stay
time for PDS. The first part of the observation, about 40\% of the total, was
characterized by a lower flux and a higher absorption --- the 6--40 keV flux
passed from 0.34 to 0.64 photons~cm$^{-2}$~sec$^{-1}$ in the two states. This
is probably due to the passage of the neutron star through clumpy circumstellar
material \cite{776}.

\section*{SPECTRAL ANALYSIS}

\begin{figure}
\vspace{7cm}
\centerline{\includegraphics{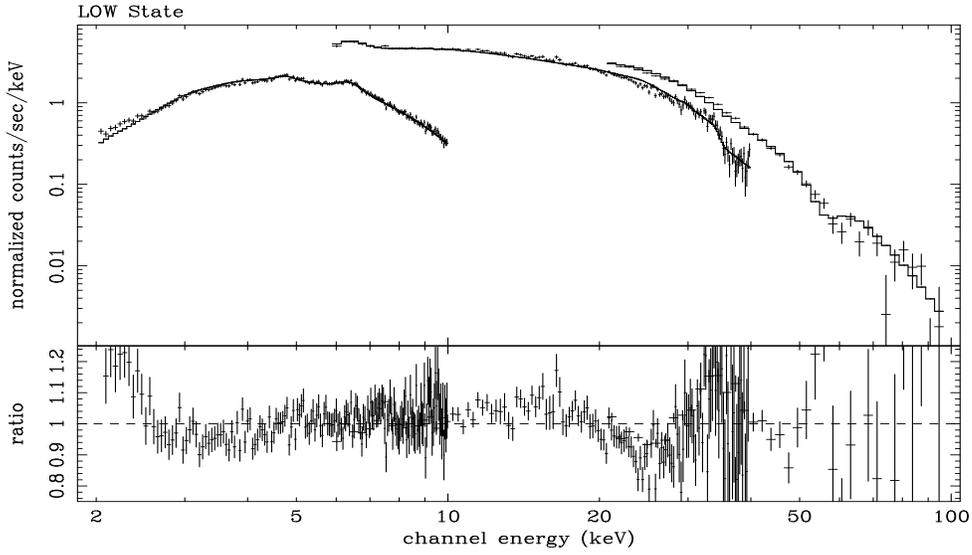}}
\caption[]{\B\ wide-band 2--100 keV spectrum of \V\ during the LOW state. The
continuum has been modelled by a power law modified by a high energy cutoff
\cite{303}. The inclusion of an Iron emission line and a CRF at $\sim 55$ keV
was not able to describe the data, in particular the flattening in the
10--30 keV range.}
\label{fig_spectra}
\end{figure}

In Fig.~\ref{fig_spectra} we show the results of the spectral fit to the 2--100
keV \V\ spectrum with the usual (for an X--ray pulsar) power law modified by a
high-energy cutoff \cite{303}. We added to the model an Iron emission line and
a CRF at $\sim 55$ keV. The spectral results are summarized in
Table~\ref{spectral_fit}, from which it is evident that this model is not able
to adequately describe the complex \V\ continuum, and in particular the
flattening of the spectrum in the 10--30 keV. A fit with {\em two\/} power laws
modified by an exponential cutoff --- the so-called NPEX model --- is able to
describe this flattening \cite{1581}, although a more detailed model is needed.
It is noteworthy that it is in this energy range that the pulse shape changes
dramatically from a five to a two peak structure.

\begin{table}
\caption[]{Fit results to the \V\ spectra for both the two intensity states.
The continuum has been modelled with a power law modified by a high energy
cutoff \cite{303}. All quoted errors represent 90\% confidence level for a
single parameter.}
\label{spectral_fit}
\begin{tabular}{p{6cm}cc}
 & \multicolumn{1}{c}{\bf LOW State} & \multicolumn{1}{c}{\bf HIGH State} \\
\noalign{\smallskip} \tableline
wabs (cm$^{-2}$) \dotfill & $7.86\pm 0.09$ & $1.63\pm 0.03$ \\
PhoIndex         \dotfill & $1.23\pm 0.07$ & $1.12\pm 0.04$ \\
cutoffE (keV)    \dotfill & $24.3\pm 0.1$  & $24.74\pm 0.07$ \\
foldE (keV)      \dotfill & $13.4\pm 0.1$  & $12.11\pm 0.06$ \\
LineE (keV)      \dotfill & $6.44\pm 0.02$ & $6.47\pm 0.02$ \\
Sigma (keV)      \dotfill & $0.21\pm 0.04$ & $0.20\pm 0.02$ \\
\noalign{\smallskip} \tableline \noalign{\smallskip}
Ecyc (keV)       \dotfill & $55.4\pm 0.5$  & $55.9\pm 0.3$ \\
Width (keV)      \dotfill & 10 (fixed)     & 10 (fixed) \\
Depth            \dotfill & 2  (fixed)     & $5.3\pm 0.5$ \\
\noalign{\smallskip} \tableline \noalign{\smallskip}
$\chi^2_{\rm dof}$ (dof) \dotfill & 2.9 (334) &  9.0 (335) \\
\noalign{\smallskip} \tableline
\end{tabular}
\end{table}

\section*{TIMING ANALYSIS}

\begin{figure}
\centerline{\epsfxsize=1.2\textwidth\epsffile{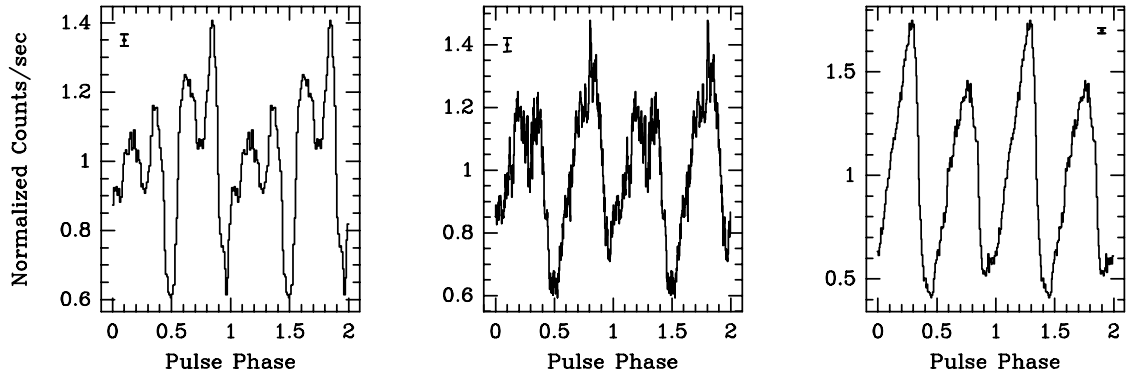}}
\caption[]{\V\ pulse profiles as observed by the three operative NFIs aboard
\B. From left to right: 2--10 keV (MECS3), 6--40 keV (HPGSPC), 20--100 keV
(PDS).}
\label{pulse}
\end{figure}

\begin{figure}
\centerline{\epsfxsize=0.55\textwidth\epsffile{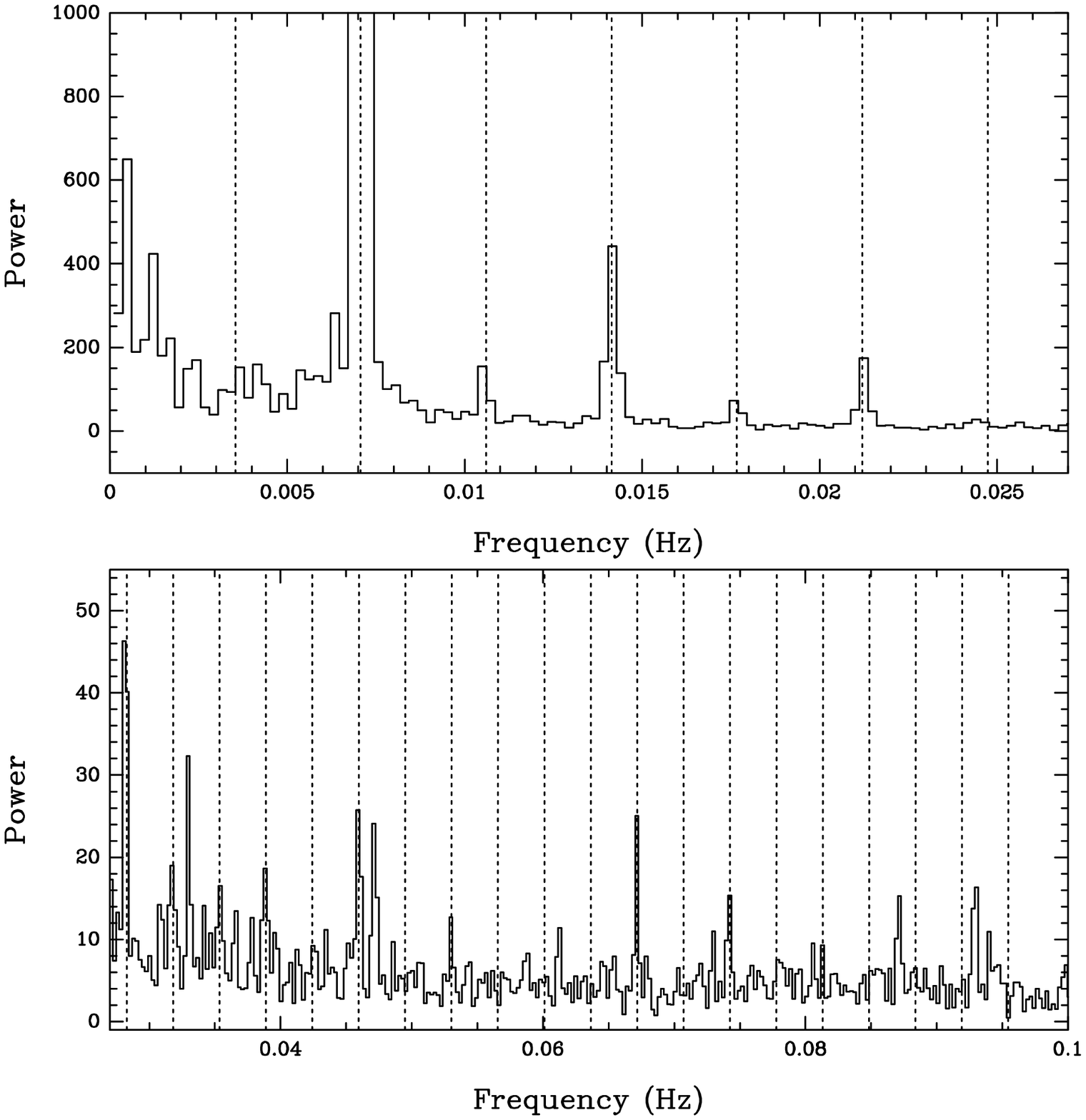}
\epsfxsize=0.55\textwidth\epsffile{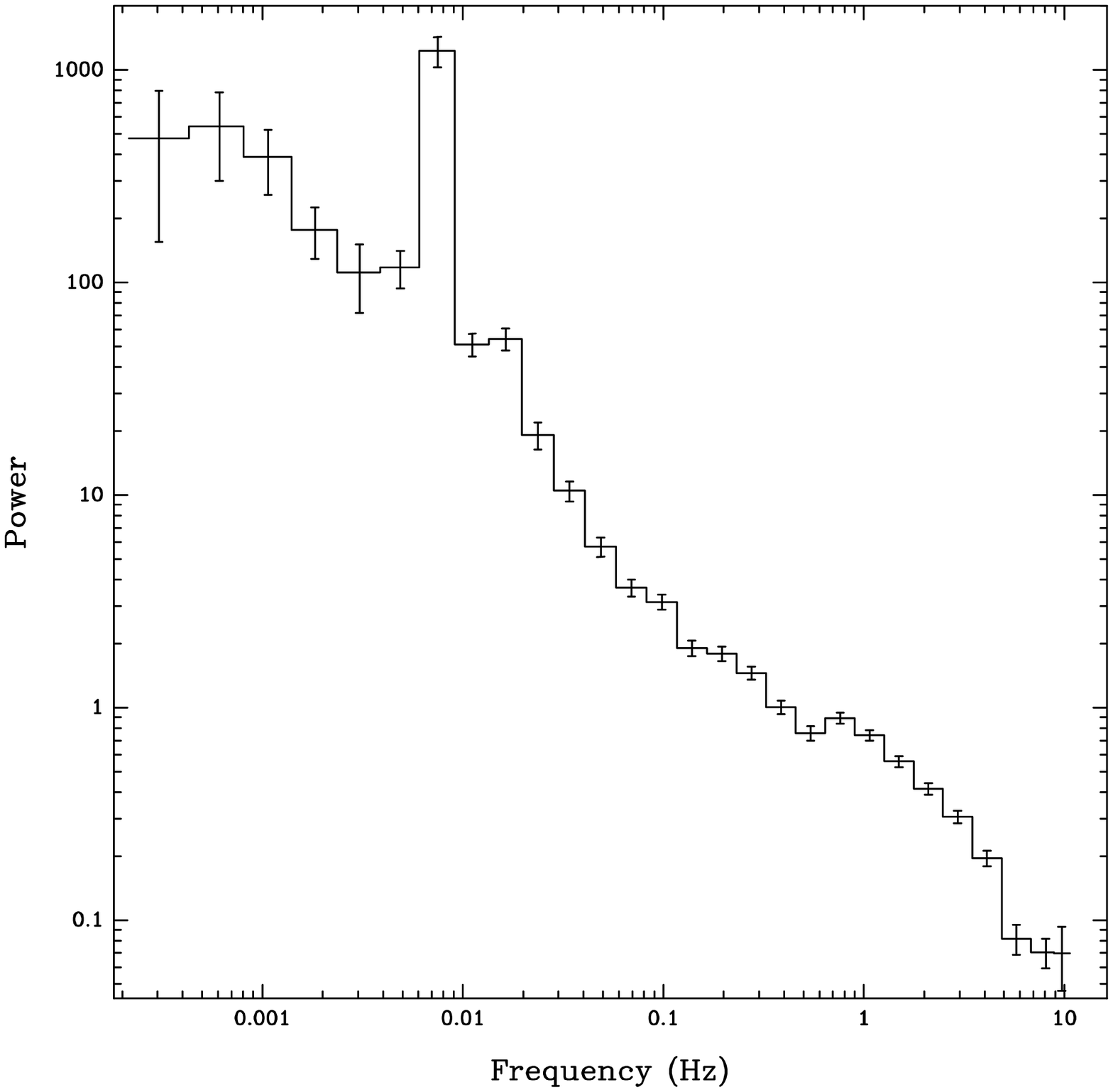}}
\caption[]{Results of the Fourier analysis performed on the PDS data. On the
left we show the power spectrum in linear scale, in order to better enhance the
effects due to the coherent component (harmonics are shown with dotted lines).
The second harmonic is the most prominent because of the double-peak structure
of the pulse profile at this energy. On the right we show the same data in
a logarithmic scale, in order to evidence the aperiodic component. Note the
{\em knee\/} at $\sim 1$ Hz.}
\label{fourier}
\end{figure}

Each photon arrival time was corrected to the solar system barycenter in order
to decouple effects due to Earth and satellite motion. We then defined a {\em
fiducial point\/} in the pulse profile --- the minimum between the two main
peaks --- and determined the arrival times corresponding to these points. Those
points were then fit with a line corrected for the Doppler delays due to the
orbital motion (we assumed the ephemeris given by \cite{705}). We find a pulse
period of $283.206 \pm 0.001$ sec for $T_o = 2,450,288.5$ JD. In
Fig.~\ref{pulse} we show the pulse profile as determined from  the three
operative NFIs instruments. It is noteworthy the dramatic change of pulse shape
from low (five peaks) to high (two peaks) energy, with the transition clearly
visible in the HPGSPC data.

We also performed a Fourier analysis on the data, and the results are shown in
Fig.~\ref{fourier} for the PDS data. From the point of view of the coherent
component, we find that the second harmonic is more pronounced than the first
because in this energy range the pulse is double-peaked. For low energy data
this behavior is reversed. For what concerns the aperiodic component, we find a
sort of {\em knee\/} at about 1 Hz, indicating a $\sim 1$ sec time scale
typical of processes that occur at the magnetospheric limit. This result is
very similar to that obtained by EXOSAT \cite{DDF}, in which a bump at $\sim 1$
Hz and a high frequency cutoff was observed.

\bigskip

{\em Acknowledgements.\/}
The authors wish to thank the \B\ Scientific Data Center staff for their
support during the observation and data analysis. This research has been funded
in part by the Italian Space Agency.

\end{document}